\documentstyle[twocolumn]{article}
\input{epsf}
\newtheorem{theo}{Theorem}

\newcommand {\beq} {\begin{equation}}
\newcommand {\enq} {\end{equation}}
\newcommand {\ra} {\rangle}
\newcommand {\la} {\langle}

\newtheorem{lemm}{Lemma}

\newtheorem{coro}{Corollary}
\newtheorem{deff}{Definition}

\newcommand{\ignore}[1]{}

\input{epsf}

\begin{document}

\title{A Quantum to Classical Phase Transition in Noisy Quantum Computers}
\author{Dorit Aharonov\thanks
{Computer Science division, UC Berkeley, E-mail: doria@cs.berkeley.edu}}

\maketitle

\begin{abstract}
The fundamental 
problem of the transition from quantum to classical physics 
is usually explained by decoherence, and viewed as 
a gradual process.  The study of entanglement, or quantum correlations, in
noisy quantum computers implies that in some cases the transition
from quantum to classical is actually a phase transition. We
define the notion of entanglement length in $d$-dimensional noisy
quantum computers, and show that a phase transition in entanglement 
occurs at a critical noise rate, where the entanglement length
transforms from infinite to finite. Above the critical noise rate, 
macroscopic classical behavior is expected, whereas below the critical 
noise rate, subsystems which are macroscopically distant one from another 
can be entangled. 

The macroscopic classical behavior in the super-critical phase 
 is shown to hold not only for quantum computers, 
but for  any quantum system composed of macroscopically many 
finite state particles,  with local 
interactions and local decoherence, subjected to some additional conditions.
 This phenomenon
 provides a possible 
explanation to the emergence of classical behavior in such systems. 
A simple formula for an upper bound on the
 entanglement length of any such system in the super-critical phase
 is given, which can be tested experimentally.   
\end{abstract}

\section{Introduction}
Quantum computation is a fascinating subject which manifests the
peculiarities of quantum mechanics and uses them in order to
achieve an advantage in terms of computational power over
classical computers. Shor's algorithm\cite{shor1} is the
most astonishing known example for such an advantage in computational
power: It enables one to factor an integer in polynomial time
using a quantum computer,  whereas the best classical algorithm
for this task is sub-exponential. The advantage of quantum
algorithms such as Shor's algorithm over classical algorithms
suggests, though not proves, that the computational complexity class
of quantum computation is not polynomially equivalent to that of
classical, even randomized, Turing machines.

In real life, quantum computers will be subjected to noise. 
We will make here the assumption that the noise is probabilistic 
and local, meaning that each particle, each time step, 
suffers a certain faulty event with probability $\eta$, which 
is referred to as the noise rate. 
Quantum computing is now known to maintain its full computational
power even in the presence of such local noise, as long as the noise 
rate is
weaker than a certain threshold $\eta_0$\cite{aharonov1,kitaev0,knill1}. 
On
the other hand, it is known\cite{aharonov2} that  when the noise
in the system is stronger than a  much higher threshold, $\eta_1$, 
the quantum computation can be simulated efficiently by a
classical Turing machine. Trying to put together the two results,
we learn that there are two regimes of noise in the quantum
computer, in which the computational power of the system is
qualitatively different. For weak noise, the computational power
is fully quantum, whereas for strong noise it becomes
classical. This raises the following question:  What is the
physical difference between the two noise regimes, which reflects
itself in the difference in computational power, and how does the
transition between the two different physical behaviors occur?

 It turns out that an answer to these questions
 can be given in terms of  the behavior of entanglement, or quantum
 correlations in the system.
 Perhaps the best way to explain the notion of entanglement
  is by saying what
 entanglement is not: We say a state in the Hilbert space of
 a composite system $A\otimes B$ is non entangled,
if two persons, Alice and Bob, each having access to one side of
the system,  could construct the overall state
  by applying local quantum
  operations on their side,
  and exchanging  classical information, by, say, talking on the
  phone.
   Any state in the composite system  $A\otimes B$ that cannot
   be constructed in this way is said to be {\it entangled}.
   Here we will be interested not only in whether or not states
   are entangled, but rather in the amount
  of entanglement in quantum  states. Several possible 
definitions for the amount of entanglement has been suggested:
The entanglement of formation\cite{bennett16}, the asymptotic entanglement 
of formation\cite{bennett16} and the asymptotic entanglement 
of distillation\cite{bennett16, rains}. 
 All these definitions are equally  suitable for the purposes
  of this paper, and the results hold for any measure 
of entanglement which satisfies certain continuity requirements. 

To study the behavior of entanglement in noisy quantum computers,
we define  the notion of  {\it entanglement length}. 
 Roughly speaking,
the entanglement length is the rate of decay of the entanglement
between two disjoint sets of particles,
 as a function of the distance between the two sets.
This is the analogous quantity to 
correlation length in statistical physics, except that here
we will be interested in correlations between two subsets of the
system, rather than in two-point correlations.
We study the behavior of the entanglement length
in the noisy quantum computer as a function of the noise rate. 
We find that there exists a noise rate, $\eta_1$, which depends on the 
geometry of the system, such that 
 the entanglement length is finite for $\eta_1<\eta\le 1$. This means 
that the entanglement between two sets of particles decays
exponentially with the distance between the 
two sets for this range of noise rates.  
On the other
hand, the entanglement length is shown to be infinite in the range
$0\le \eta< \eta_0$, where $\eta_0$ is the threshold for 
fault tolerance. 
This means that  the entanglement between two
sets of particles is independent of the distance between the two
sets. These two facts show the existence of a phase transition in
entanglement at a non-trivial noise rate $\eta_0\le\eta_c\le\eta_1$.
 The
system in the sub-critical regime behaves
quantumly even on the macroscopic scale- two sets of particles,
within macroscopic distance, can share a lot of entanglement, so
there is long range entanglement in the system.   In the
super-critical phase, where entanglement decays exponentially with
the distance, the system behaves classically on the macroscopic scale, 
because two macroscopic  subsets within macroscopic distance 
 are practically non entangled.

The results here are by no means 
specific to quantum computers. 
In fact, we show that macroscopic classical behavior 
is expected above the critical noise rate for any
 macroscopic quantum system
with local interactions, and local noise,
where we make the additional assumption that 
there is time separation between two interactions
in which  one particle participates.  
This shows that  a phase transition in entanglement is expected in any 
such quantum system which exhibits long range 
entanglement in the absence of noise.   
Moreover, our results can be verified 
experimentally in any such system, as long as the density matrices 
of subsystems can be measured accurately enough.   
The entanglement length can then be numerically 
computed (this is a difficult computational task, 
but possible for small subsystems)
 and compared  to the finite entanglement length which is predicted 
by our analysis.

The emergence of 
classical macroscopic behavior in large quantum systems
has been an intriguing area of research for the last several decades.
   Perhaps the most common and acceptable explanation so far 
is by decoherence, i.e.  interactions with the environment 
which cause the system to lose its quantum features. 
See, for example, 
\cite{zurek1,decoherence} and references therein. 
 This explanation, however, predicts  a gradual transition 
from quantum to classical behavior. 
 The most interesting implication of the results presented 
in this paper
 is that they suggest an alternative way to explain the
transition from quantum macroscopic behavior
 to classical macroscopic behavior
in certain physical systems, which is qualitatively different 
from the standard gradual explanation.
The origin for the abrupt phase transition from quantum  to
classical   predicted 
by our results is that in our model, we combine  
the decoherence process with the assumption that noise, as well 
as interactions, are local, where the behavior we are interested in 
is global.
\ignore{ The importance of the results here is that 
it seems that our assumptions are applicable in many general 
cases. }

The first part of the proof involves showing that the entanglement
decays exponentially with the distance when $\eta$ is larger than
a certain threshold. To do this, we use a method due to Aharonov
and Ben-Or\cite{aharonov2} to present the density matrix by
mixtures of clustered states, and then we study the behavior of
the sizes of the clusters, evolving in time, using a mapping of
the problem to percolation. Known results from percolation 
theory\cite{grimmett} imply
an
exponentially decaying bound on the probability for distant sets
of particles to be connected in the same cluster, and this implies
exponentially small entanglement between the two sets. For the
second part of the proof, i.e. in order to show that the
entanglement length is infinite for weak noise, we use fault
tolerant quantum computation, which enables one to create long
range entangled states in the noisy quantum computer. 

We start by
defining the notion of entanglement, and entanglement length, 
and then proceed to prove the strong noise case and the weak noise case. 
We conclude with several open questions regarding possible 
implications to the transition from quantum to classical.   

\section{Entanglement}
The notion of {\it entanglement} is associated with 
a state of a quantum system 
composed of two systems, $A$ and $B$. 
The term entanglement refers to the quantum correlations between 
$A$ and $B$, in a state which lives in $A\otimes B$.
Remarkably, two parts of a composite quantum  system 
can exhibit very strong 
correlations which cannot be explained  classically, 
unless one drops a very important axiom in physics, namely locality. 
The remarkable phenomena  which can be exhibited due to entanglement
between two quantum systems  
were first discovered by Einstein, Podolski and Rosen\cite{epr}, more 
than $60$ years ago, and manifested in Bell's inequality more than 
$30$ years ago\cite{bell,bell1}. 
However, the elusive phenomena of entanglement is still 
far from being understood. 
 
In this paper we will be interested in  
 the {\it amount} of entanglement
 in quantum systems; we therefore need a good measure 
of entanglement. 
One very important requirement on such a measure is that 
the entanglement in any state cannot increase by classical 
communication and local quantum operations on the $A$ and $B$ sides 
separately; this is the whole 
essence of the term entanglement. 
We will denote such a process involving 
 local operations 
 and classical communication by 
 LOCC.  A good measure of entanglement should also be additive.

A natural way to construct a measure of entanglement  is to ask 
whether there is an elementary unit of 
entanglement, so that any state can be constructed with LOCC 
given sufficiently many such entanglement units. 
It turns out that there exists exactly such a unit: 
the Bell state, 
$
\frac{1}{\sqrt{2}}(|0\ra\otimes|0\ra+|1\ra\otimes
|1\ra)$. It was shown\cite{bennett15,bennett16}
 that any  bipartite quantum state 
can be generated by Alice and Bob using only LOCC operations   
given that sufficiently many Bell states are a priori shared between 
$A$ and $B$. One can try to use the number
of elementary units required to construct a state 
as a good measure of the entanglement in this state.  
It is reasonable to take the asymptotic limit of such a process,
and to define the entanglement in a state as the following limit.
Let $\phi$ be our quantum state, and  
let $k_n$ be  the number of Bell states  required to 
generate a state, $\phi_n$, and let $\phi_n$ approach the state 
$\phi^{\otimes n}$ as $n$ tends to infinity. 
Bennett {\it et al}\cite{bennett16}
  defined the infimum over all such processes,  of $k_n/n$,  as $n$ tends to 
infinity, 
as the asymptotic entanglement of formation of the state $\phi$.
Let us denote this measure by $E_f^\infty$.  
This measure is clearly additive, and can also be shown not to 
increase by LOCC\cite{bennett16}.

An equally natural definition would be the converse one, called
the asymptotic entanglement of distillation, in which one is interested 
in generating as many Bell states as possible by applying LOCC 
on many copies of the original state $\phi$. 
The asymptotic limit of the ratio between the number of Bell
states generated in this way,  and $n$, was defined 
in \cite{bennett16} to be the asymptotic  entanglement of distillation. 
A more rigorous definition was given by Rains\cite{rains}.
Let us denote this measure by $E_d^\infty$.

Fortunately, for pure states these two measures 
coincide, and have a very beautiful form. 
As was shown in \cite{bennett15},
they are exactly the  von-Neumann entropy of 
the reduced density matrix on one part 
of the system.
\beq E(A:B,|\phi\ra\la\phi|)=S(|\phi\ra\la\phi||_A).\enq
The entropy of entanglement thus possesses both additivity 
and monotonicity under LOCC, and also 
 behaves nicely in many other ways.

The situation for mixed states, however, is much more interesting. 
It turns out that though the asymptotic distillable entanglement and the 
asymptotic entanglement of formation coincide on pure states, 
 there are very interesting differences between them
when mixed states are considered. Clearly, 
the asymptotic entanglement of distillation 
is not larger than  the asymptotic entanglement of formation\cite{bennett16}.
The question of whether there exist states in which 
$E_f^\infty$ is strictly larger than $E_d^\infty$ is still open. 
This irreversible process, in which not all 
of the entanglement which was inserted into the state 
can be distilled, is called   
{\it bound entanglement}\cite{horodecki} and is now being extensively  
studied. \ignore{(eg. \cite{bound})}

The asymptotic  entanglement of formation  is believed to
 be equal to the following quantity, 
 called the {\it entanglement 
of formation}\cite{bennett16}, and denoted by $E_f$. 
$E_f(\rho)$ is the least expected entropy 
of entanglement of any ensemble of pure states 
realizing  $\rho$, or more formally: 
 \begin{equation}\label{ef} E_f(A:B,\rho)=\min_{\sum_i w_i|\alpha_i\ra\la
\alpha_i|=\rho} \sum_i w_i E(A:B,
|\alpha_i\ra\la\alpha_i|).\end{equation} 
The question of whether $E_f$ is equal or not to $E_f^\infty$ 
depends on whether $E_f$ is additive. It is believed that 
indeed it is the case that $E_f$ is additive, but this is not known.

Let us survey what is known about the above three entanglement
 measures, in terms of  convexity and continuity.
Entanglement of formation is trivially convex. 
 Asymptotic entanglement of formation can also
 be shown to be convex, using the law of large numbers. 
Currently it  is not known whether asymptotic distillable 
entanglement is convex or not. 
As for continuity, the situation is even less clear.
It is known that the above three entanglement measures are 
continuous, in the 
sense that if a sequence of density matrices 
$\sigma_n$ converges to a density matrix $\rho$ in the trace metric, 
then the entanglement in $\sigma_n$ converges to the entanglement 
in $\rho$: 
\beq
\lim_{n \longrightarrow \infty} \sigma_n =\rho
~~~ \Longrightarrow ~~~
\lim_{n \longrightarrow \infty} E(\sigma_n) = E(\rho).
\enq
However, we will be interested in how different can the entanglement 
of two very close density matrices be. Entanglement 
of formation was recently shown\cite{neilsen}
 to have very strong 
continuity properties, in the following sense: 
 Given two density matrices of a bipartite Hilbert space 
of dimension $d\times d'$,   
which are within $\epsilon$ distance one from another,
in the trace metric, the entanglement of formation of the two matrices 
is at most $\epsilon$ times some linear function in 
 $\log(d)$ and $\log(d')$, 
plus a term independent of $d$ and $d'$ which goes to $0$ 
as $\epsilon$ goes to $0$:
\begin{eqnarray}
&&|E_f(A:B|\rho)-E_f(A:B|\sigma)|\le  \\\nonumber
&& 9|\rho-\sigma|\log(\max\{d,d'\})- 
|\rho-\sigma|\log(|\rho-\sigma|)). \end{eqnarray}
This strong continuity implies that when two density matrices 
of $n$ finite state particles are polynomially close one to another
 (in the number of particles), the entanglement 
of formation between them is also polynomially small.  
It is not yet known whether the asymptotic measures 
of entanglement possess these nice continuity properties, or not.

In this paper we work with the entanglement of formation, $E_f$, which is 
known to be both convex and strongly continuous. 
However, it should be stressed that the phenomena 
presented in this paper depend very weakly on the exact properties 
of the  measure of entanglement which is being
used.  
The  results in this paper hold, 
with straight forward modifications,  for any  measure of entanglement  $E$
which is  continuous in a sufficiently strong sense, meaning that 
two density matrices which are $\epsilon$ apart have entanglements 
not different by more than $\epsilon$ times some polynomial
 in the number of particles.

\section{The Model of the Quantum System}\label{model}
We are interested in quantum systems composed of 
$n$ two-state particles, embedded on a $d$-dimensional
lattice. 
Such quantum particles are usually called {\it qubits}
in the context of quantum computers. 
The Hilbert space of $n$ such particles is the tensor product 
of $n$  two dimensional complex vector space, ${\cal C}^2$, where  
the basis of ${\cal C}^2$
 is standardly taken to be $|0\ra$ and $|1\ra$. 
The system is initialized with a certain state (usually a tensor 
product state, but not necessarily) and evolves in time 
via interactions between the particles.  
Time is discretized into time steps 
and all interactions are assumed to be instantaneous and occur 
at integer times. 
In this model, particles interact only with their  nearest neighbors
on the lattice.   An important assumption 
is that one particle cannot interact with more than 
one other particle at a time. 
For simplicity, we will assume that the particles interact alternately with
particles to each of their sides.   For one dimension, i.e. an
array of particles, this means that a particle interacts with a
particle to its left and to its right alternately. 
The interaction
graph can be easily viewed  in a $d+1$ dimensional scheme, which
for $d=1$ looks as follows:

{~}

\begin{picture}(150,80)(-30,20)
\put(10,20){\vector(0,1){95}} \put(9,15){\shortstack{$a$}}
\put(24,15){\shortstack{$b$}} \put(39,15){\shortstack{$c$}}
\put(25,20){\vector(0,1){95}}\put(40,20){\vector(0,1){95}}
\put(55,20){\vector(0,1){95}} \put(70,20){\vector(0,1){95}}
\put(85,20){\vector(0,1){95}} \put(100,20){\vector(0,1){95}}
\put(130,20){\vector(0,1){95}} \put(140,68){\shortstack{$Time$}}
\put(130,20){\circle*{2}} \put(134,20){\shortstack{$0$}}
\put(134,30){\shortstack{$1$}} \put(134,40){\shortstack{$2$}}
\put(134,50){\shortstack{$3$}} \put(134,60){\shortstack{$4$}}
\put(134,70){\shortstack{$5$}} \put(134,80){\shortstack{$6$}}
\put(134,90){\shortstack{$7$}}\put(134,100){\shortstack{$8$}}
\put(130,30){\circle*{2}} \put(130,40){\circle*{2}}
\put(130,50){\circle*{2}} \put(130,60){\circle*{2}}
\put(130,70){\circle*{2}} \put(130,80){\circle*{2}}
\put(130,90){\circle*{2}}\put(130,100){\circle*{2}}

\put(10,30){\line(1,0){15}} \put(40,30){\line(1,0){15}}
\put(70,30){\line(1,0){15}} \put(10,50){\line(1,0){15}}
\put(40,50){\line(1,0){15}} \put(70,50){\line(1,0){15}}
\put(10,70){\line(1,0){15}} \put(40,70){\line(1,0){15}}
\put(70,70){\line(1,0){15}} \put(10,90){\line(1,0){15}}
\put(40,90){\line(1,0){15}} \put(70,90){\line(1,0){15}}
\put(25,40){\line(1,0){15}} \put(55,40){\line(1,0){15}}
\put(85,40){\line(1,0){15}} \put(25,60){\line(1,0){15}}
\put(55,60){\line(1,0){15}} \put(85,60){\line(1,0){15}}
\put(25,80){\line(1,0){15}} \put(55,80){\line(1,0){15}}
\put(85,80){\line(1,0){15}} \put(25,100){\line(1,0){15}}
\put(55,100){\line(1,0){15}} \put(85,100){\line(1,0){15}}

\end{picture}

\begin{figure}[h!]
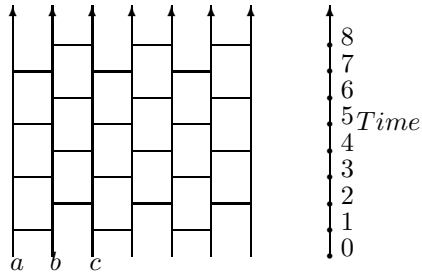

\begin{center}
\caption{The vertical axis corresponds to time, and the horizontal
axis corresponds to space. Horizontal edges connect two
interacting particles. Particles interact alternately with
particles to their left or to their right. }\label{lattice}
\end{center}
\end{figure}

Two particles can interact via an arbitrary interaction, and we do
not assume anything about the nature or strength of each interaction.
 After the
interactions are turned off, and before the next step of
interactions is turned on, we apply the noise step.
The noise is assumed here to be local and stochastic, 
 meaning that each particle with a certain probability $\eta$ 
undergoes an arbitrary fault process. 
As was shown in \cite{aharonov1, gottesman} such $d$-dimensional 
noisy quantum circuits 
are capable of performing fault tolerant quantum computation,
as long as the noise rate is smaller than a certain threshold. 
The threshold, however, is worse than the threshold without the nearest
neighbor restriction, by one or two orders of magnitude, depending
on the dimension.

We make here another assumption, and restrict the noise to be 
one of the following two processes.
The first process, namely independent stochastic
collapses, is a process in which at each time step, each particle is
measured with independent probability $\eta$, in a fixed but arbitrary
basis. Alternatively, we can use  the depolarization model, in
which at each time step, each particle,
 with independent probability $\eta$, is replaced by a particle in a
completely mixed state. 
  In the rest of the paper, we will assume that the noise  model
is independent stochastic collapses, but all results can
be easily stated using stochastic
depolarization.

It should be noted that the results of the paper hold 
when relaxing several of the assumptions we have made. 
The results apply to particles with any finite number of states,  
not necessarily qubits, 
and different particles need not have the same number of possible states.   
The exact form of the alternating interactions is not important,
since any interaction graph of nearest neighbor interactions 
is a subgraph of the alternating graph, if time is scaled by a 
factor of $2d$.   
The assumption of instantaneous interactions is also not essential, 
as long as the interactions last for less than a time step, 
so that there is a time interval in which decoherence takes place when 
the particle is not participating in any interaction.   
The results hold also in the case of noisy interactions, 
with some noise rate $\delta$.
We will see that the proof of the upper bound 
on the entanglement length, for large $\eta$,  
holds regardless of the amount of
 noise in the interactions, $\delta$, because 
the proof only uses the noise occurring between interactions.  
The proof of the lower bound on the entanglement length, 
for small $\eta$, goes through as long as $\eta+\delta$ is smaller than 
the threshold for fault tolerance. Hence, 
it is straight forward to include in our model the above generalizations.  
For simplicity, however, we will work with the model defined above, 
of 
two state particles with  noiseless instantaneous
interactions.

\section{Entanglement Length}
For quantum systems which are embedded on a lattice,   
the notion of distance between sets of particles is well defined. 
In this case, one can define the {\it entanglement length} 
in the system. We would like to define an analogous quantity to 
the standard correlation length from statistical physics. 
In this case, one says that the correlation
 length in a physical system is $\xi$ if
the correlation between the outcomes of a certain 
observable $O$ measured at sites $a$ and $b$, decays
exponentially with the distance between them, $d(a,b)$, where the distance
is scaled in the decay factor by $\xi$: \beq
<O_aO_b>\propto e^{-\frac{d(a,b)}{\xi}}.\enq 
 More precisely, $\xi$ is defined to be the following 
quantity: 
\beq\label{limit}
\xi^{-1} =\lim_{d(a,b)\longmapsto \infty} \left(\frac{-log<O_aO_b>}{d(a,b)}\right).
\enq

In analogy with correlation length, we could have defined 
  the entanglement length in the quantum system to be
 $\mu$ if
the entanglement between two particles, $a$ and $b$, decays
exponentially with the distance between them, $d(a,b)$, where the distance
is scaled in the decay factor by $\mu$: \beq\label{expdec0}
E(a:b)\propto e^{-\frac{d(a,b)}{\mu}}.\enq
\ignore{Or 
 \beq\label{expdec1}
\mu^{-1} =\lim_{n\longmapsto \infty} \left(\frac{-log(E(a:b)}{d(a,b)}\right).
\enq}

However, there are a few problems with this definition, which 
will force us to modify it slightly. 
 The first modification is necessary due to the fact that 
entanglement is a non-local quantity. 
It might well be that the system contains a  lot of entanglement, 
but small subsets of the system are completely unentangled. 
For example, in fault tolerant quantum computers, 
  the entanglement is bound to be shared by
large sets of qubits, and in order to see entanglement it is
necessary to probe large subsets of the system. 
 We will therefore be interested not in two point correlations, but
in entanglement between two sets $A$ and $B$ of arbitrary sizes.

Another problem  is the following. In systems 
which are homogeneous in space and time, one can easily take the
limit of the size of the system to infinity, and therefore the
asymptotic behavior in equation (\ref{limit}) is well defined.
 However, we are interested in 
fault tolerant quantum
computers, which are not homogeneous in space, nor in time.  
Roughly speaking, we will say that the entanglement length in 
the system is $\mu$ if 
the entanglement between any two sets $A$ and $B$ 
 is {\it bounded} by a function which decays  exponentially
 with the distance between the sets, where the decay factor is scaled by 
$\mu$. The fact that we are interested in  a bound, and not in exact 
behavior of any pair of sets, allows for non homogeneity in space. 
To allow for  non homogeneity in time, 
we will consider the average 
 entanglement between $A$ and $B$ over the time 
from $t=0$ to $t=\infty$.  
This corresponds to the following behavior: 
 \beq\label{expdec3}
<E(A:B)>_{t=0}^{\infty}\propto poly(|A|,|B|)e^{-\frac{d(A,B)}{\mu}}.\enq
where $|A|$ is the number of particles in $A$ and similarly for $B$. 
We allow the  additional polynomial factor due to the fact that 
for sets which are not too large compared to the distance
between them, the exponential decay dominates the 
polynomial in the sizes of $A$ and $B$, and what we will see is merely 
an exponential decaying behavior.  We claim that it 
is not reasonable to consider
two sets of particles which are very large, and to study the 
behavior of the entanglement they share as a function of 
the distance between them, in the range where 
that distance is extremely small compared to the sizes of the sets. 

The characterization of $\mu$ by equation (\ref{expdec3})
is very helpful to keep in mind. 
We can also make the definition of entanglement length  more rigorous, 
by  giving it a similar form to that of equation (\ref{limit}).
This would be useful when one actually wants to 
 calculate the entanglement length.  
\ignore{ 
In order to take the limit in equation 
(\ref{expdec1}), we need sets  $A$ and $B$ with distance 
 $d(A,B)$ which goes to infinity. 
However, in the non-homogeneous case, this might depend on the 
way these limits are taken. 
Therefore, we
need to be more careful in explaining what we mean by
asymptotic behavior. 
To generalize the homogeneous case,} 
In order to do that, we first
 need to make the notion of  a quantum system more 
precise. 
In the non-homogeneous case, 
it is not clear what the notion of an infinite system 
means.  We therefore define a quantum (infinite) system to be
a sequence of quantum systems, $Q_n$, where $Q_n$ consists of 
 $n$ particles. We think of $n$ as growing to $\infty$, but 
for a given $n$ $Q_n$ is a finite system in space, which evolves 
 in time from 
$t=0$ to $t=\infty$. 
Since each $Q_n$ is finite in space, in order to take a limit  
 similar to that of equation (\ref{limit}), 
we need to consider 
a sequence of pairs of sets,    $A$ and $B$
which belong to larger and larger systems. 
We thus add a subscript $n$ to the subsets $A_n$ and $B_n$, 
indicating that they belong   to the quantum system $Q_n$. 
We would now like to translate 
the fact that we are interested in sets 
which are not too large compared with their distance 
to a precise restriction on the  sequences of sets  
 $\{A_n\}$,$\{B_n\}$. The weakest condition
 which we can impose, to avoid pathologic cases,
 is that  
$\lim_{n\longmapsto \infty} |A_n|\cdot |B_n|/exp(d(A_n,B_n)) =0$, 
 meaning that 
the sizes of  $A_n$ and $B_n$ are not growing 
exponentially or faster than exponentially with the distance between them. 
Finally, we want to take care of the fact that we are interested in 
the largest entanglement  length which can be observed 
in the system. This  corresponds to taking the  
 infimum over all such sequences of $A_n$ and $B_n$.  
All this translates to the following  definition: 
\begin{deff}\label{expdec4}
The entanglement length $\mu$ of a quantum system $\{Q_n\}_{n=1}^\infty$ is defined by: 
\[
\mu^{-1} =\inf_{\{A_n\}_{n=1}^{\infty},\{B_n\}_{n=1}^{\infty}} 
\liminf_{n\longmapsto \infty} \left(\frac{-log<E(A_n:B_n)>_t}{d(A_n,B_n)}\right)
\]
where for all $n$,   $A_n$ and $B_n$ are disjoint sets 
 in $Q_n$, and the sequences $\{A_n\}_{n=1}^{\infty},\{B_n\}_{n=1}^{\infty}$
satisfy  
$\lim_{n\longmapsto \infty} |A_n|\cdot |B_n|/exp(d(A_n,B_n)) =0$.
\end{deff}
Note that if one  plugs into
 definition  (\ref{expdec4}) the exponential behavior 
of equation (\ref{expdec3}), 
 the contribution 
of the polynomial factor in equation (\ref{expdec3}) 
tends to zero due to the requirements on $A_n$ and $B_n$, 
and the correct $\mu$ pops out. 
Though definition (\ref{expdec4}) might seem complicated, 
calculating the above infimum 
turns out to be very simple in all our applications.

\ignore{ 
-----------------------

Thus, our modified definition of entanglement length 
is 
\beq\label{expdec2}
\mu^{-1} =\inf_{a,b} \lim_{n\longmapsto \infty} \left(\frac{-log(E(a:b))}{d(a,b)}\right).
\enq

 Now, a subtle question should be clarified. 
The question is whether one should take the infimum over 
all possible disjoint sets, or restrict the sets which are considered. 
We claim that  in reasonable scenarios, one will 
be interested in cases in which the distance between 
the sets $A$ and $B$ is of the order of their sizes, or at least 
not exponentially smaller than their sizes.

Now, it is no longer reasonable to consider an exponential 
decay multiplied by a constant, because the maximal entanglement 
$E(A,B)$ is not a constant anymore if we allow arbitrarily large 
sets $A$ and $B$.   It is more reasonable 
  to scale the entanglement with  
 the maximal amount of entanglement which can exist between 
 $A$ and $B$, which  is linear in the number of particles 
 in $A$, and $B$ (for finite state particles.) 
  In fact,
we will use a slightly rougher definition, in which 
we allow the coefficient to 
depend polynomially on the sizes of $A$ and $B$.
This is reasonable to do since if the distance 
between $A$ and $B$ is of the order of their sizes, or larger, 
the exponential decay will dominate any polynomial in $|A|$ and $|B|$.
  We will therefore change equation \ref{expdec0}
to the following:

 More precisely one
is interested in the asymptotic behavior, and therefore we take
the limit: \beq \xi=\lim_{d(a,b)\longmapsto
\infty}-\frac{d(a,b)}{\log(<c_ac_b>).\enq}

The last problem with the definition 
 of entanglement length according to 
equation \ref{expdec1} is that it depends strongly on the entanglement 
length of the  initial state 
of the system. 
For example, consider the trivial system which applies no interactions, 
 where the entanglement between $A$ and $B$ in the initial state 
is of the order of $n$, the number of particles in the system. 
This
 entanglement is very quickly  lost, due to collapses, and no new
 entanglement 
is generated in the system since there are no interactions. 
The system therefore 
relaxes to its typical unentangled state very quickly.   
However, this fact does not show in the average entanglement in the system. 
In fact,  the average entanglement from $0$ to $t$ remains  significantly 
larger than $0$ for very long time intervals- $t$ should
 be of the order of the 
size of the system, $n$, for the effect of the initial entanglement 
on the average entanglement to vanish.   
This strong dependence on initial conditions is undesirable. 
We want to say that a system has finite entanglement length 
even if its initial condition has long entanglement 
length, as long as the system 
relaxes to its typical behavior of short correlations quickly. 
Since it is reasonable to assume that the time over which a system is 
observed is polynomial in its size, we will say that negligible effects are those which disappear exponentially fast. 
Therefore
we add a time dependent correction term to 
equation \ref{expdec0}:
\beq\label{expdec}
E(A:B)\propto poly(|A|,|B|)e^{-\frac{d(a,b)}{\xi}}+poly(|A|,|B|)
e^{-\frac{t}{\xi_t}}\enq
 The correction term 
corresponds to a logarithmic relaxation time to the
behavior described by the exponential decay in space.
Alternatively, one can simply start considering the quantum system 
after $log(n)$ time steps, 
i.e. after it has relaxed to its typical entanglement length, or restrict 
the possible initial conditions to unentangled states. 
These variants of the definition do not  
affect the decision of whether or not  
a system does or does not have finite entanglement length. }


\section{Clustered Density Matrices}\label{sim}
We now proceed to study the entanglement length 
in $d$-dimensional noisy quantum circuits, in the 
 strong noise regime. In this case, 
we will try to bound the entanglement in the system from above.
 Now a very useful
observation is in place. After a particle was hit by the noise
process, it is no longer entangled with the rest of the system.  In
other words, in both noise processes which we consider,
 the density matrix after
applying the noise process can be written in the form \beq {\cal
E}\rho= (1-\eta)\rho+\eta \sum_i p_i\rho^Q_i\otimes \rho_i^q \enq
where the index $q$ refers to the noisy particle, and $Q$ refers to
the rest of the system. For example, for the stochastic noise process, 
in which the last qubit is measured in the basis $\{|0\ra,|1\ra\}$, 
the resulting density matrix would be of the form:
\beq {\cal
E}\rho= (1-\eta)\rho+\eta \sum_{i=0}^{1} Pr(i)\rho^Q_i\otimes |i\ra\la i| \enq
where $\rho^Q_i$ is the density matrix of all but the last qubit,
under the condition
 that the last qubit is measured to be in the state $|i\ra$. 
  We use this observation as follows. We
will aim to present the density matrix in such a way that lack of
entanglement translates to tensor product structure. In other
words, we will present a density matrix as a mixture of tensor
product states, as follows:
 \begin{eqnarray}\label{clusterdef} \rho(t)&=&\sum_i w_i \rho_i(t),\\\nonumber
  \rho_i(t)&=&\rho^1_i(t)\otimes\cdots \rho_i^{m_i}(t)\end{eqnarray}
where $\rho^j_i(t)$ is a density matrix which describes a set of
particles $A_i^j$, and for each $i$ the sets $A_i^j$ are a
partition of the system. 
These sets of supposedly entangled
particles are called clusters. 
It should be understood here that given 
a density matrix, there is no single way to present it as a mixture 
of clustered states. However, we will define the representation according 
to the dynamics of the process which generated $\rho$, so that our 
representation will be well defined.

Our goal would be to find a way to represent the matrix 
as a mixture of clustered states with as small clusters as possible.
The intuition is that we want to give an upper bound on the amount of 
entanglement in the system. 
When all the clusters are of size one,
there is no entanglement in the system. We will see later, that
this can be generalized to say that small localized clusters imply no
entanglement between distant sets. We will thus try to keep 
the clusters as small as possible. 
The way we do this is as follows. 
In a quantum computer, the initial state is a basic state, which
is a pure state in which all qubits are in tensor product with one
another: \beq\rho=\rho(1)\otimes \rho(2)\otimes \cdots \otimes
\rho(n).\enq Thus, for $t=0$, all clusters are of size $1$. Given
any clustered states description of the density matrix at time
$t$, we can obtain a clustered state description for the matrix at
time $t+1$ as follows.
 From each participant in the
mixture, $\rho_i(t)$, we obtain $\rho_i(t+1)$ which will be a
mixture of clustered states.  $\rho(t+1)$ will then be a mixture
of all $\rho_i(t+1)$. To obtain $\rho_i(t+1)$ from $\rho_i(t)$, we
first apply the interaction step, and then apply the noise step.

To apply the interaction step of time $t$, we apply for each
interaction at that time step the unitary transformation
corresponding to the interaction, on the appropriate pair of
particles. If the two particles are from one cluster, then we
simply apply the appropriate unitary matrix, corresponding to the
interaction, on the density matrix describing this cluster, and
there is no need to change the clusters. However, if the two
particles are from two different clusters, we can no longer keep
the two clusters in tensor product, because in general they will
be entangled.  Therefore, we first join the two clusters together,
by taking the tensor product of the density matrices describing
the two clusters, and then apply the appropriate unitary matrix on
the new big cluster. The resulting state after all interactions of
time step $t$ were applied is therefore, in general, a clustered
state with larger clusters than the state $\rho_i(t)$.

 We then apply the noise step on the
resulting clustered state. Recall that a
measurement detaches a particle from its cluster, and thus after a
measurement the particle is a cluster of its own.  To apply the
noise process, we transform the state to a mixture of states,
which are the results of all possible combinations of which
particles where measured, with the appropriate probabilities.
Clusters in the state can only shrink due to this process.

We would now like to understand the typical size of clusters in
this representation of the density matrix. Before we do that in a
more formal way, let us gain some intuition.  If the system were
noise free,  very soon all the clusters would become one giant
cluster of $n$ particles.  What makes the situation more
interesting  are the stochastic collapses, which separate a
measured particle from its cluster, thereby decreasing the size of
the cluster by one, and creating another cluster of size one. One
can view the noisy quantum evolution in time as a struggle between
two forces: The interactions, which tend to join clusters and
entangle the different parts of the system,  and the stochastic
collapses, which tend to detach particles from their clusters,
thereby  destroying this entanglement constantly. A crucial point
here is that the two competing forces are matched in power, since
they both operate on a linear number of particles $\theta(n)$ each
time step. We thus expect a 
critical error rate, $\eta_c$,
 at which the two forces are equal, and at which some transition
 between the dominance of the entangling interaction process
 transforms to the dominance of the disentangling noise process.
 We now go on to see this phenomenon more rigorously,
 using a map
 to a percolation process.


\section{The Percolation Process}\label{perc}\label{percpt}
It turns out that the dynamics of the clusters in the above
description
 are intimately connected with a percolation process
on the {\em quantum circuit} itself. The percolation process on
the graph is defined as follows: For each time step, each vertical
edge, along the $i'$th wire, between  time $t$ and $t+1$, is
erased with independent probability $\eta$. In the cluster
picture, this corresponds to the collapse of the $i$'th particle
between time steps $t$ and $t+1$, which disentangles the
particle's past and future. Thus an event in the probability space
of the noise process, i.e. a specific choice of which particles
collapsed during the process,  is mapped to an event in the
percolation process, in which the corresponding vertical edges are
erased. Since events in the stochastic noise process correspond to
members in the mixed density matrix, 
 we have a map between
clustered states arising from our cluster dynamics, 
 and realizations of the percolation process.
 This map preserves the probability measure.

  We now claim that clusters
in the clustered state correspond to connected components in the
percolation process:

\begin{lemm}\label{corres} {\bf Correspondence Lemma: }
Two particles $a$ and $b$ are in the same cluster at time $t$, in
one realization of the noise process in the cluster model, iff
$(a,t)$ and $(b,t)$ are connected
 in the corresponding realization of the percolation model.
\end{lemm}

{\bf Proof:} To prove this combinatorial  lemma  
we use induction on $t$. 
  For the base of the induction, $t=0$, the correspondence 
is true by definition. Let us now assume that the lemma is correct 
for $t$, and prove for $t+1$. 
To apply the induction step, the following 
 observation comes
in handy. Each path that connects $(a,t+1)$ and
$(b,t+1)$ in the percolation process, 
is actually a concatenation
of alternating paths, occurring either after time $t$ or at 
times up to $t$. \ignore{Clearly the transition from one path to the 
next path in the concatenation always occurs 
at time $t$. }
We denote the points at which  the different concatenated paths
connect one to another  
 by
$(x_1,t_1),...,(x_{2k},t_{2k}).$ It is easy to see that there is always 
an even number of such points, and that $t_1=t_2=....=t_{2k}=t$.
Let us call the particles $x_1,...,x_{2k}$ the connection particles. 
We shall also denote $a=x_0$, $b=x_{2k+1}$. 
A  schematic example for the one dimensional quantum circuit case 
is shown in figure $2$. 

\begin{figure}[h!]
\centerline{\vbox{\epsfxsize=2in\epsfbox{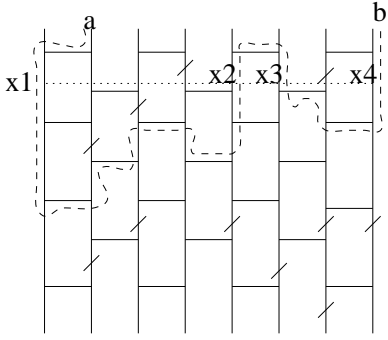}}}
\caption{A path
connecting the two particles, $a$ and $b$,  at time $t+1$
 can be represented as a
concatenation of paths which are restricted alternately to 
the time intervals  $[0,t]$ and $[t,t+1]$.  
The path $(a,t+1)\longmapsto (b,t+1)$ in the figure is a concatenation of the 
paths  $(a,t+1)\longmapsto (x_1,t)\longmapsto (x_2,t) 
\longmapsto (x_3,t)\longmapsto (x_4,t) \longmapsto (b,t+1).$
}
\end{figure}

Let us now prove the first direction: 
 Let $a$ and $b$ be two
particles connected at time $t+1$ in the percolation model. 
We want to show that $a$ and $b$ are in the same cluster at time 
$t+1$. 
To see this, we show that all particles $a=x_0,x_1,..,x_{2k},x_{2k+1}=b$ 
are in the same cluster at time $t+1$.  
For the pairs of particles $x_{2i+1}$ and
$x_{2i+2}$, i.e. pairs in which the first particle has odd 
index,  this is true since they 
are connected by a path confined to time steps $t$
 or earlier, so by the induction assumption, they 
are in the same cluster at time $t$. Moreover, 
 none of the connection particles 
 collapsed between time steps $t$ and
$t+1$, due to the fact that they connect between   
a path 
before time $t$ and a path after time $t$ (or vice-versa).
Therefore  $x_{2i+1}$ and $x_{2i+2}$
 are in the same cluster also at time $t+1$. 
(In the schematic example, this shows that particles 
$x_1$ 
and $x_2$ are in the same cluster at time $t+1$,
 and similarly $x_3$ and $x_4$.) 
Now by definition of the connection particle, 
there is a path after time $t$ 
connecting $x_{2i}$ and $x_{2i+1}$, which means that   
 $x_{2i}$ and $x_{2i+1}$ interact at time $t+1$. At the edges 
of the chain, i.e. for $i=0$ or $i=k$,
it might be that  $x_{2i}$ and $x_{2i+1}$ are
 the same particle, and therefore they are trivially connected. 
(In the example, this corresponds to the 
interaction at time $t+1$  
between $a$ and $x_1$, and between $x_2$ and $x_3$, 
 and to the fact that $b=x_4$. ) 
By the definition of the clusters' evolution in time,
the fact that particles interact imply that 
 their clusters are joined, and therefore
 $x_{2i}$ and $x_{2i+1}$ are in the same 
cluster at time $t+1$.\ignore{ (In the example, this corresponds to the fact that 
the particles $x_2$ and $x_3$ are in the same cluster at time $t+1$, 
and similarly the pair $a$ and $x_1$.   $b$ and $x_4$ 
are trivially in the same cluster since they are the same particle.) } 
Combining this with the fact that $x_{2i+1}$ and $x_{2i+2}$ are connected at $t+1$,  this implies that all the particles  $a=x_0,x_1,..,x_{2k},x_{2k+1}=b$ 
are in the same cluster at time $t+1$, which
 completes one direction of the induction step.

Let us now prove the other direction of the induction step. 
We want to prove that there is a path connecting 
$(a,t+1)$ and $(b,t+1)$ in the percolation process, assuming that
 $a$ and $b$ are in the same cluster at time $t+1$. 
This cluster of time step $t+1$, which contains $a$ and $b$, 
was generated by joining together several smaller 
clusters, which existed after the noise step of
 time step $t$.  It is easy to see that  
 there is a subset of those clusters, 
$ C_{-1}, C_0,...,C_{k-1},C_{k}$, such that $a\in C_{-1}$ and $b\in C_{k}$,
 and such that 
each two subsequent clusters, $C_i$ and $C_{i+1}$,
were connected at time $t+1$ by a unitary gate. 
Note that each cluster $C_i$, except maybe 
$C_{-1}$ and $C_{k}$, consists of at least two particles: one particle, 
denoted by $x_{2i+1}$,  
 participates in an interaction with a particle from the preceding 
cluster in the chain,  $C_{i-1}$, 
and the other particle, $x_{2i+2}$,  participated in an  interaction 
with a particle from the next cluster in the chain, $C_{i+1}$. 
To construct a path from $a$ to $b$ at time $t+1$, we first 
note that by the induction assumption, there is 
 a path connecting  the two particles 
in the same cluster, $x_{2i+1}$ and  $x_{2i+2}$, at time $t$.   
Moreover, these particles did not collapse between time step 
$t$ and $t+1$, because if they did collapse, they would have
belonged to a cluster consisting of one particle exactly.
Hence there is a path connecting them at time $t+1$. 
A path between the second particle of $C_i$ and the first particle 
of $C_{i+1}$ exists because by definition they interact at time $t+1$. 
Paths from $a$ to the first particle $x_1$, and from 
the last particle $x_{2k}$ to $b$, exist either because 
they interact at time $t$, or they are simply the same particle. 
This enables us to construct a concatenated path from $a$ to $b$ 
 at time $t+1$. $\Box$

\ignore{
to this cluster by interactions at time $t+1$.  
The fact that $a$ and $b$ are in the same cluster at time $t+1$, 
implies  that there exists a set of particles
$X_1,...X_{2k}$ such that the following conditions are satisfied
\begin{enumerate}
\item At time $t$ $X_{2i+1}$  is in a different cluster than $X_{2i+2}$ and
$X_{2i}$ is in the same cluster as $X_{2i+1}$.
\item At time $t+1$ unitary matrices are applied on particles $X_{2i+1}$ and
$X_{2i+2}$.
\item  $X_i$ did not collapse between the time steps
$t$ and $t+1$.
\item  (A unitary matrix is applied on $a$ and $X_1$ at $t+1$)

or

($X_1$ is in the same cluster as $a$ at time $t$ and $a$ did not
collapse between $t$ and $t+1$)
\item  (A unitary matrix is applied on $b$ and $X_{2k}$ at $t+1$)

 or

($X_{2k}$ is in the same cluster as $b$ at time $t$ and $b$ did
not collapse between $t$ and $t+1$)
\end{enumerate}
 The claim is that there is a path connecting $a$ and $b$ in $B_{t+1}$.
We build this path as a concatenated path from $a$ to $X_1$ to
$X_2$... to $b$, where the paths between $X_{2i}$ and  $X_{2i+1}$
exist due to the induction assumption, and the paths between
 $X_{2i+1}$ and  $X_{2i+2}$ exist because of conditions $2$ and $3$.
The edges of the path (From $a$ to  $X_1$  and from  $X_{2k}$ to
$b$) exist because of conditions $4$ and $5$.}

We can now investigate the sizes of the connected components in
the percolation model and then translate our findings to the
cluster model. Such percolation processes are known to exhibit a
phase transition at a critical point, below which there is a
connected component of linear size, but
 above which the typical connected
components is of logarithmic size. Moreover, the connected
components in the super-critical phase  are localized, in the
sense that the probability for two particles  to be in the same
connected component decays exponentially with the distance between
these particles.

Let us first release the restriction to a percolation on a square
of size $n \times T$ and consider the infinite lattice.  
We show the existence of a phase transition in
 connectivity for percolation on
 the infinite lattice, and from this we will be
able to get information about the finite case. One more
simplification which is useful is to notice that by contracting
each edge which corresponds to an interaction to one point, we do
not change the connectivity properties of the process, 
and the resulting percolation process is  
the usual model of percolation.\ignore{ on  translational invariant lattices, 
in which each edge is erased with independent probability 
$\eta$. 
This simplifies the original process, 
in which there were two types of edges: 
Interaction edges which were always present, and 
time evolution edges which were erased with certain probability.}  
For example, in the two dimensional lattice associated with the
 one dimensional quantum circuit, the interaction edges 
are exactly the horizontal edges in figure $1$, and so  
after contracting each of these edges to one point, 
the resulting percolation process   
is  the standard percolation on the square lattice, which is rotated 
by $45$ degrees. 
The contraction therefore transforms the problem to standard 
 bond percolation on translational invariant lattices, 
which is pretty well understood.

In bond percolation on translational invariant lattices, 
 one usually uses $p$ as the probability for one edge
to be present, so in our case $p=1-\eta$. One can define the
critical $p$,   $p_c$, to be the smallest probability in which the
point $0$ belongs to an infinite connected component with positive
probability. In translational invariant lattices, this is also equal
to the smallest $p$ in which the expected size of $0'$s connected component
 is infinite\cite{mms,ab,chayes}
\begin{eqnarray}
p_c&=&inf\{p~|~
Pr_p(|H(0)|=\infty)>0\}=\\\nonumber
&=&inf\{p~|~E_p(|H(0)|)=\infty\}.
\end{eqnarray}
where $H(0)$ is the connected component of $0$.

A theorem by
Hammersley\cite{chayes,grimmett}, asserts that for
translational invariant lattices, for $p<p_c$, the probability
$\tau(x,y:p)$ for $x$ to be connected to $y$, decays exponentially
with the distance:
\begin{equation}\label{decay}
  \tau(x,y:p)\le \exp\left(-\frac{d(x,y)}{\xi(p)}\right)
\end{equation}
with  $\xi(p)<\infty.$ Above $p_c$, the probability for  $0$ to be
connected to infinity is larger than $0$ by definition of $p_c$, so 
$\xi(p)=\infty$ for $p>p_c$. 

Let us denote by $p_c(d+1)$ the critical $p_c$ for bond percolation 
on the lattice corresponding to a quantum circuit of dimension $d$. 
For one dimensional quantum systems, it is easy to calculate the 
critical probability $p_c(1+1)$. After contracting the interaction edges, we
simply get percolation on $Z^2$, for which it is known\cite{grimmett} that
the critical probability is a half. Hence: 
\beq p_c(1+1)=\frac{1}{2}.\enq
 For higher dimensions, we can bound $p_c(d+1)$ away from $0$ and $1$, 
but the bounds are not tight:
\begin{lemm}
\( \frac{1}{3}\le p_c(d+1)\le\frac{1}{2^{1/d}}.\)
\end{lemm}
{\bf Proof:}
The upper bound comes from the fact that 
  the projection of the $d+1$
percolation with parameter $p$ on  a $1+1$ percolation gives
percolation in $1+1$ with  parameter $p^{d}$. This is true since after a
particle interacts with a particle along one axis, it waits $d$
time steps before it interacts again with a neighbor along the
same axis.  This gives the upper bound,  since if the original process had
exponentially decaying correlations, it cannot be that in the
projected process we are above the phase transition where $0$ is
connected with constant probability to infinity.
The lower bound is derived from a standard argument 
which reduces the problem to a branching process. 
A branching process is a process which starts with one node, 
and each nodes gives birth to $k$ nodes with some probability 
distribution 
$p(k)$, independent of the other nodes. 
It is a standard result (See \cite{feller} , for example)
that when the expected number 
of descendents for each node is less than $1$, 
the dynasty dies in finite time with probability $1$. 
To construct the corresponding branching process, 
observe that  the degree of the interaction 
graph, after contracting the horizontal edges, 
is exactly $4$, regardless of $d$. 
Starting with the point $0$, we regard each of  its neighbors to which 
it is connected in the percolation   as an 
  ancestor of a dynasty. The descendents of each such node 
are all its neighbors,
except for $0$, in the percolation process; 
Each descendent has its own descendents, and so on.  
When we encounter a node which is already in the dynasty, we do not count it
again; this way we have a tree. 
Clearly, if the branching process is finite, then the connected component 
of $0$ in the percolation process is definitely finite. 
However, each neighbor of $0$ is the ancestor of a dynasty 
in which the expected number of descendents of each node 
is exactly the number of its neighbors which are not yet in the dynasty, 
times $p$. Since the degree of the graph is $4$, and one of the 
neighbors is the node's ancestor, the expected number 
of descendents is at most $3p$, which is less than $1$ for $p$ less than 
$\frac{1}{3}$. This gives the desired result. $\Box$

The analysis of the percolation process has taught us that 
when $p$ is smaller than the critical point for percolation, $p_c$,  
 the connected components in the system are small.  
Going back to the density matrices, using the correspondence lemma, 
 this implies 
 that for $\eta < 1-p_c$, the clusters are small and localized.  
This will be used in the next section 
to prove an upper on the entanglement between 
distant sets in this noise regime. 

\section{Finite Entanglement Length}
From the correspondence to percolation, 
we have that for $\eta < 1-p_c$,  the density
matrix of the quantum system can be approximated by 
a mixture of clustered states with
localized clusters of logarithmic size.
Thus, distant subsets of
particles are with high probability contained in non-intersecting
clusters, i.e. most of the weight of the density matrix is
concentrated on states in which there is no entanglement between
the two subsets. 
The weight of the states in which there
is entanglement between the two sets decays exponentially with the distance
between the sets. By continuity of entanglement, this implies that 
the entanglement between the two sets decays exponentially with the distance. 

We can now show,
that the entanglement between any two sets of
particles decays exponentially with the distance between them, 
when  the noise rate $\eta$ is such that $1-\eta$ is sub-critical 
in the percolation process. 
The rate of the decay is the {\it entanglement length} of the
system. 
The entanglement between the two sets becomes negligible already when the 
distance is of the order of $\log(n)$ particles.
 This translates to the following theorem: 

\begin{theo}\label{theo1}
Consider a $d$ dimensional quantum circuit with
 nearest neighbor interactions, subjected to local noise 
of the type of stochastic depolarization 
or stochastic collapses, with noise rate $\eta$.
If the circuit is initialized with an unentangled state, i.e. 
a tensor product state, and  if $\eta>1-p_c(d+1)$,
then  the entanglement of formation 
between any two sets of qubits $A$ and $B$ at any time $t\ge 0$ decays
exponentially with the distance between the two sets:
\[ E_f(A:B)\le \min\{|A|,|B|\}|A|\cdot|B|e^{-\frac{d(A,B)}{\xi(1-\eta)}}.\]
For a general initial state, a similar formula is true 
except for a correction term which decays exponentially with time:  
\[ E_f(A:B,\rho(t))\le \min\{|A|,|B|\}\left(|A|\cdot|B|e^{-\frac{d(A,B)}{\xi(1-\eta)}}+\right.\]\[~~~~~~~~~~~~\left.+ n\min\{|A|,|B|\}e^{-\frac{t}{\xi(1-\eta)}}\right).\]
\end{theo}

\noindent{\bf Proof:} Let us start with the simple case, in which the initial state 
is a complete tensor product, i.e. all clusters are of size $1$. 
By equation \ref{decay},
 the probability for two particles $A$ and $B$ to be connected
decays exponentially in the distance between them.
 The correspondence lemma (\ref{corres}) implies that the
probability for  two particles to be in the same cluster at time
$t$ is equal to the probability they are connected in the
percolation model at time $t$. i.e. the probability for two
particles  from $A$ and $B$ to be in the same cluster is bounded
above by $\exp(-d(A,B)/\xi(1-\eta))$. Thus, the probability for any
pair of particles from $A$ and $B$ to be in the same cluster is
bounded above by $|A|\cdot|B|\exp(-d(A,B)/\xi(1-\eta))$. The density
matrix can thus be written as a mixture of one density matrix with
weight smaller than
 $|A|\cdot|B|\exp(-d(A,B)/\xi(1-\eta))$, and another density matrix which
is a mixture of density matrices, where in all these matrices all
the particles in $A$ are in different clusters than all the
particles in $B$. The reduced density matrix
 to $A,B$ of the
second matrix is thus separable, and contains no entanglement
between $A$ and $B$. 
By convexity of entanglement of formation, 
the entanglement in the entire density matrix
is bounded above by the entanglement in the first
density matrix, times the weight of this matrix. The entanglement
of the first matrix is at most 
 the number of qubits in the system, and this gives the desired result.
(For measures of entanglement which are not convex,
but strongly continuous, one should replace the term $\min\{|A|,|B|\}$
 by the appropriate polynomial from the continuity bound.)

We now proceed to the general initial state. 
We will give an upper bound for the case in which the initial 
state is one big cluster, and any other case is trivially 
implied by it. 
To do this, we have to understand where we have 
 used the fact that the initial state is not entangled. 
This was used 
  for the base of the 
induction in the correspondence lemma, 
 where the fact that all clusters are of one qubit
corresponds to the fact that in the percolation graph,
 the initial connected components at
 time $0$ are all of size $1$.  
 To adapt the situation to the case in which 
 all particles are in one big cluster
at time $0$,   we add a horizontal line of length $n$ 
 connecting all particles to
one big connected component at time $t=0$. The correspondence lemma 
then goes through. However, equation \ref{decay} no longer holds. 
To correct it, we add to it a term which corresponds to 
the probability for $A$ to be connected to $B$ by a path 
that goes through time $t=0$, i.e. through the additional new line 
we have added to the graph. 
For such a path to exist, both  $A$ and $B$  need
 to be connected to time $0$.  
The probability for any one of the qubits in $A$ to be connected to any 
one 
of the $n$ qubits at time $0$ is at most 
$n|A|$ times the probability for one qubit 
at time $t$ to be connected to one qubit at time $0$, which is at most 
$exp(-\frac{t}{\xi(1-\eta)})$ by equation \ref{decay}. The same argument 
applies for the connection from $B$ to time $0$, and this 
 gives the desired result. 
$\Box$

This shows that the system cannot create entanglement between far
sets of particles: Roughly speaking, the typical range of
entanglement is microscopic.
This is true for any initial condition, where 
the relaxation time to the typical unentangled state 
is of the order of  $\log(n)$ steps.

This result implies an upper bound on the 
entanglement length in the quantum system above the critical noise rate,
 and in particular  shows that it is finite.
This is done by simply taking the limit in the definition of entanglement length (\ref{expdec4}), which gives:
  \begin{coro}\label{coro1}
The entanglement length $\mu(\eta)$ of a $d$ dimensional quantum circuit with
 nearest neighbor interactions, subjected to local noise 
of the type of stochastic depolarization 
or stochastic collapses, with noise rate $\eta$, satisfies:
\[\mu(\eta)\le \xi(1-\eta)\]
and in particular $\mu(\eta)$ is finite for   
 $\eta>1-p_c(d+1)$. 
\end{coro}

This gives a bound on the entanglement length, in terms 
of the correlation length in classical bond percolation. 
The correlation length of a given  lattice 
can be easily estimated by computer experiments, and  
  analytical bounds are given 
in \cite{chayes, grimmett}. 

\section{Infinite Entanglement Length}
We now want to concentrate on the other noise regime, and  show
that below the critical noise the entanglement length is infinite.
One might naively think that this can be deduced from the fact
that the density matrix is a mixture of clustered states with
linear sized clusters. However, there is a difficulty in pursuing
the connection between clusters and entanglement for this purpose,
because of the following reason. The density matrix is actually a
mixture of many clustered states. The mixture of two clustered
states, with very large clusters, can be a density matrix in which
the clusters are of size one. One example of such a case is a
mixture of the two states, $\frac{1}{\sqrt{2}}(|0^n\ra+|1^n\ra)$ and $\frac{1}{\sqrt{2}}(|0^n\ra-|1^n\ra)$,
the mixture of which is a non entangled state. Thus, the sizes of
the clusters can be used for upper-bounds on entanglement, but it
is not clear how to use them in order to show a lower bound on the
entanglement in the system.

We therefore need to use different techniques for 
lower bounds on entanglement. We will use techniques from 
quantum computation. 
A quantum computer embedded on a lattice is a special case of the
quantum systems we are discussing. The particles are quantum bits,
and the interactions are fixed according  to the algorithm.
Therefore, corollary \ref{coro1} shows 
 that the entanglement length is finite above
the critical noise rate also in fault tolerant quantum computers.
For fault tolerant quantum 
computers we can also analyze the other side of
the noise scale, and  show that the entanglement length  in the
system is infinite if the noise rate is below a certain threshold.
  We will use the 
 threshold result\cite{aharonov1,kitaev0,knill1} for
 fault tolerant quantum computation, which shows that
 quantum computation can be made robust to noise, using quantum error
 correcting codes, as long as the noise is smaller than a certain threshold.
In fact, here we need the slightly stronger version of the
threshold  result\cite{aharonov1,gottesman}, which asserts that
this can be done even when the quantum system is embedded on a $d$
dimensional lattice. The threshold is then $\eta_0(d)$, which 
for $d=1$ is estimated to be $10^{-7}$\cite{aharonov1}. 
 In the fault tolerant range, two distant sets
of qubits can be entangled, and remain entangled for a long time, 
 with the amount of entanglement independent of the distance.

We now give an example of a quantum computer which exhibits
entanglement among far parts of the system when $\eta<\eta_0(d)$,
but the entanglement length is finite for noise
 $\eta>1-p_c(d+1)$. The idea is that
a fault tolerant computer can simulate  any quantum state,
including states which contain entanglement between sets of qubits
which are far apart. Hence, we will
 construct a quantum algorithm in which there is entanglement
between two far parts of the system, and make it fault tolerant.
This can be done in many ways, but here is a simple example, 
for $d=1$. 
Divide the set of qubits to three sets, $A,B,C.$ We will create
entanglement among $A$ and $B$, while leaving the qubits in the middle,
 $C$, in
a basic state. This will be done by constructing the state:
\beq\label{state}
\frac{1}{\sqrt{2}}\left(
|0^m\ra_A\otimes |0^n\ra_C\otimes |0^q\ra_B+
|1^m\ra_A\otimes |0^n\ra_C\otimes |1^q\ra_B\right)
\enq
on a fault tolerant quantum computer, and keeping this state for a
long time, by applying error corrections.
This state indeed contains entanglement between the two registers 
$A$ and $B$, which are $n$ sites apart. 
 The algorithm which constructs such a state
is very simple, and uses only two basic quantum gates: 
The Hadamard gate, which is a one qubit gate applying the following unitary 
transformation \begin{eqnarray} |0\ra &\longmapsto&
\frac{1}{\sqrt{2}}( |0\ra+|1\ra)\\\nonumber
|1\ra &\longmapsto&
 \frac{1}{\sqrt{2}}(|0\ra-|1\ra),\end{eqnarray}
and the controlled NOT gate 
 which is a two qubit gate applying the following
unitary transformation 
(said to be applied {\it from} the first qubit to the second one):
 \beq |a\ra \otimes |b\ra  \longmapsto |a\ra\otimes 
|a\oplus b\ra,\enq
where $\oplus$ means addition mod $2$.  
Using these gates, it is easy to create the state \beq
\frac{1}{\sqrt{2}}(|0^{m+q}\ra+|1^{m+q}\ra) \enq on the first $m+q$ qubits, by
applying a Hadamard gate on the first qubit and then controlled NOT gates
 from
the first qubit to the second, from the second to the third, and
so on. Then, we want to swap the $m+1,...,m+q$ qubits to register
$C$. To do this, we first swap the last qubit in $B$ with qubits 
to its right until it gets to the last site in $C$; 
In the same way we bring the one before last qubit in $B$ to the one
before last site in $C$, and so on until all qubits in $B$ are 
in the right most sites of $C$, which achieves the desired state
with only nearest neighbor interactions.

This algorithm by itself is not fault tolerant, and in the
presence of any amount of noise, i.e. $\eta>0$, the entanglement
in the system will be lost immediately. However, we can make this
algorithm fault tolerant by the methods in \cite{aharonov1,kitaev0,knill1}, 
as long as $\eta$ is smaller than $\eta(d)$, 
 the threshold for fault tolerance 
for $d$-dimensional quantum computers\cite{aharonov1,gottesman}. 
These results are too complicated to explain here in details. 
In a nutshell, fault tolerance is 
achieved  by encoding the qubits using  
 quantum error correcting
codes, and computing the algorithm on the encoded states, while
applying quantum error correction on the state frequently. 
 Each qubit is replaced by
polylog($n$) qubits, encoding its state. 
The state $|0\ra$ is encoded by the state $|S_0\ra$ of polylog($n$)
qubits, and similarly  $|1\ra$ is encoded by the state $|S_1\ra$.
 Let us denote by 
 $A'$, $B'$ and $C'$ the qubits encoding 
the original sets of qubits $A$, $B$, and $C$,  respectively.
If no fault occurs, at the end of the
algorithm the state of the system will be in the 
state (\ref{state}) encoded by the quantum error correcting code:
\beq\label{encstate}
|S_0^m\ra_{A'}\otimes |S_0^n\ra_{C'}\otimes |S_0^q\ra_{B'}+
|S_1^m\ra_{A'}\otimes |S_0^n\ra_{C'}\otimes |S_1^q\ra_{B'}
\enq
normalized by a factor of $\frac{1}{\sqrt{2}}$. 
The entanglement in this state will remain there for ever if errors 
do not occur.  However, errors do occur. By fault tolerance, 
 this means that at the end of the computation
the density matrix  is polynomially close to a density matrix $\rho$ 
which can be corrected to the 
 correct state (\ref{encstate}) by noiseless quantum error corrections. 
Due to continuity of entanglement, it suffices  to argue 
that such $\rho$ contains a constant amount of entanglement. 
But this is true since we know that $\rho$ can be corrected 
to the state (\ref{encstate}) by local operations not involving interactions 
between $A'$ and $B'$. 
 Since  entanglement cannot increase by 
local operations, the entanglement between $A'$ and $B'$ in $\rho$ 
is at least as that in the state encoding (\ref{encstate}), i.e. one 
entanglement unit. 
The distance between the actual density matrix
and a correctable density matrix $\rho$ is,
 by \cite{aharonov1},  at most the number of time steps $t$ 
divided by a polynomial in $n$. This distance 
is smaller than some constant $\epsilon$ 
as long as  the number of time steps is polynomial in the size
 of the system $n$. Thus, by strong continuity
the entanglement between $A'$ and $B'$ 
 will remain bounded from below by a constant for 
 polynomially many time steps. 
After polynomially many steps, we can replace all qubits by 
qubits in the state $|0\ra$, and run the whole algorithm again. 
The average entanglement over time from $0$ to $\infty$
 is very close to one, 
since the time it takes to construct the state is much smaller than 
the polynomial time for which the entanglement remains in the system. 
 This proves the existence of a non trivial sub-critical side  of
the phase transition:
\begin{theo}
The entanglement length in the $d$ dimensional 
 fault tolerant quantum computer
defined above satisfies 
\[ \mu(\eta)=\infty\]
for $\eta$ smaller than the threshold for fault tolerance in $d$ 
dimensional quantum computers, i.e. $\eta<\eta_0(d)$. $\Box$
\end{theo}

\section{Other Quantum Systems}\label{gen}
The model of a noisy quantum computer actually holds not only for 
quantum systems designed to serve as computational devices,
 but for a  much broader class of
physical systems as well. 
We first claim that putting aside the noise process, 
any $d$ dimensional quantum system, in which the particles 
are located in space with low enough density, and 
in which interactions occur only between particles which are not 
too far apart,   
can be modeled by a quantum circuit. 
This can be done by discretizing the medium to very small cells, 
such that each cell contains at most one particle. 
Time will be discretized to sufficiently small intervals 
such  that a particle can only move 
to a neighbor cell in one time step. Then, the
movement of particles can be modeled 
by an interaction between an occupied and an un-occupied cell, and
since the density of particles is low, one particle never interacts 
with more than one other particle at the same time, so the notion of quantum
 gate is appropriate. 

The noise model which is used in this paper is quite general as well, 
when low density or instantaneous interactions systems are considered.   
During the time interval in which a particle does not participate
in any interaction, stochastic collapses are actually equivalent to
a process of local  
 decoherence. 
 Assume that each particle interacts with its own 
independent thermal bath, and the
Markovian assumption is applied, so that the environment of each
particle is renewed each time step. This corresponds to the process 
in which the   off diagonal terms in the density matrix of 
one particle decay by some factor between two interactions\cite{palma}. 
If the decoherence process operates for time         
$\Delta t$, the $(i,j)$ element of the density matrix, 
 transforms to  \beq \rho_{i,j}
\longmapsto \rho_{i,j}\exp(-\gamma\Delta t(1-\delta_{i,j}))\enq
If we set $\exp(-\gamma\Delta t)=1-\eta$, we get 
\beq \rho_{i,j}\longmapsto (1-\eta)\rho_{i,j}+
\eta\rho_{i,j}\delta_{i,j},\enq which is equivalent to a
measurement with probability $\eta$. 
 Similarly, the depolarization process can be
presented as a gradual change of the density matrix of one particle. 

The above arguments show that the model of noisy 
quantum circuits which we are discussing 
is interesting as a representative of 
the class of quantum systems with
 macroscopically many 
finite state particles, local instantaneous
 interactions and local decoherence noise. 
The analysis done in this paper regarding 
upper bounds on the entanglement length in the super-critical phase 
goes through, and therefore theorem \ref{theo1} and corollary \ref{coro1}
can be generalized to this case. 

In such quantum systems, our analysis provides an explanation 
to  the emergence of macroscopic classical behavior above the critical 
noise rate, as will be discussed in the conclusions.

\section{Experimental Verification}
Unfortunately, we do not yet have a physical realization 
of a quantum computer of more than several qubits, on which 
the existence of a phase transition in entanglement length 
in fault tolerant quantum computer  can be verified.
However, for the more general case of 
quantum systems with local interactions and local noise, 
satisfying the requirements of section \ref{gen}, 
  the bound on entanglement length in the 
super-critical regime can in principle be experimentally  testable. 
 What is needed to perform such an experiment is to be able to measure 
with high enough accuracy the joint density matrix of 
two subsets of the system. 
The entanglement between the two sub-systems can then be numerically 
approximated, using a (doable, but extremely difficult) 
minimization over equation \ref{ef}. 
The entanglement can then be found as a function  
of the distance between the sets, from which the entanglement length 
can be deduced.  
An extremely  interesting open problem is  to give a concrete design 
for such an 
experiment, for an existing quantum system, 
 and to compare the outcomes
with the entanglement length predicted by the percolation process.   

\section{Quantum-Classical Transition}\label{quantclass}
 The results presented here suggest that 
the emergence of classical macroscopic phenomena 
in large quantum systems can be attributed, in certain cases, 
to the fact that the noise rate is larger than a certain critical 
point, so that the entanglement length is finite.
However, we merely introduced in this paper a new phenomena. 
The list of questions which remain open is extremely large, 
and varies on different physical fields.

The first and most basic question should be
how general these results are. 
In this paper, we have been able to show the existence 
of a non trivial sub-critical phase in fault tolerant 
quantum computers. Are there more natural quantum systems,  
in particular systems which are homogeneous (or periodic) in
 space and in time, which have local interactions and local 
noise, which are 
 able to maintain long range entanglement in the presence of  
weak or zero local noise for a long time?    
Does a random quantum  system, i.e. 
 in which  random interactions are applied, exhibit long range 
quantum correlations?  
Such systems will provide more examples for quantum systems 
in which a phase transition in entanglement length occurs.

It should be noted here that the notion of zero noise rate does not 
trivially coincide with that of  zero temperature, 
i.e. long range entanglement 
in the ground state of the Hamiltonian of the system. 
An important observation is that
 the model discussed here
deals with non-equilibrium quantum systems.
 The quantum systems we consider here can be in a
steady state, but the density matrix is not in the Gibbs
distribution of the eigenvectors of some Hamiltonian. 
The reason for this is that we did not allow the 
system to approach equilibrium. Our noise model,  or the
interaction with the thermal bath, is limited to local
interactions. 
 This is a crucial ingredient that causes the phase
transition. It is the fact that two forces of even power
compete: local interactions in the system against local
interactions with the environment, that gives the critical noise
rate. The fact that the quantum computer does not 
achieve equilibrium despite the noise is explained by the fact 
that the system is cooled constantly by 
quantum error corrections. 
It is left as an open problem to further investigate possible 
 equilibrium phase transitions in entanglement, and the connection 
to the non-equilibrium phase transition presented here. 
The reader is referred to \cite{marro} and references therein 
for an introduction to non-equilibrium phase transitions. 
  We view the fact that we did not allow the system to achieve
equilibrium as very important in the derivation here.

\ignore{This relates to a remark regarding the 
difference between our explanation of the emergence 
of classical behavior, and the standard explanation. 
The suggestion to explain the transition from
 quantum to classical macroscopic behavior as a phase transition
 stands in contradiction to the
standard point of view of gradual transition, usually explained by
decoherence\cite{zurek1} or by taking $\hbar$ to $0$\cite{}.}

Indeed, it is worthwhile to ask which of our assumptions
regarding the properties of the physical system are
essential, and which are technical. Intuitively, the locality of both
noise and interactions seems crucial. This is true since 
this locality is what makes  the two competing 
forces, the interactions which tend  to 
entangle the system and the noise which tends  to disentangle it, 
comparable in power, which gives rise to the phase transition.   
 It seems, however, that the assumption on
the exact form of the noise process might  not be so important, 
and neither is the discretization of the
interactions. 
An important open problem  is to relax the assumptions 
used in this paper, and to generalize 
the results presented here to other noise models, and
 to the continuous case.   In particular, it is not clear how to 
generalize the results to the case in which
the particle interacts simultaneously with all its neighbors
and the environment. It seems that a considerably different 
approach would be needed in this case.

If the phenomena of the phase transition in entanglement 
is indeed general, its effect on 
our understanding the transition from quantum to classical 
physics needs to be deeply understood.    
One important question is whether there exists 
some classical or quasi-classical description of the behavior 
of a quantum  system in its  super-critical phase. 
Another, related, question is whether the existence of a
 phase transition in entanglement induces other 
quantum phase transitions at the same critical point 
in the same system.

\ignore{ Why is the
transition from quantum to classical connected at all to the
entanglement length? We claim that when the entanglement length is
finite, the system will behave classically when observed in the
macroscopic scale. This is true because two macroscopic subsets 
of the system,
which are within macroscopic distance one from another, are almost
non entangled, so there will be no quantum correlations between 
the two sub-systems.}

\ignore{It is easy to construct an homogeneous quantum system 
in which the phase transition occurs at exactly $0$. 
An example for such a system is a system in which 
random Hadamard transforms and random controlled NOT's 
are applied at each time step.}

A set of  open questions regarding the phase transition 
comes from statistical physics. For example, 
what are the critical exponents related to this phase transition? 
What is the universality class of this phase transition? 
In fact,  it is not clear that there is only one critical 
point here. In the case of the quantum 
computer, or other quantum systems,  there might well be 
an  intermediate
regime of noise, in between the two thresholds, for which the
entanglement behaves in a different way, i.e. its dependence on the
distance is neither an exponential decay nor constant. The question
of showing that there is only one critical point at which a transition
occurs from exponential decay to independence on the distance
remains open.

A very interesting problem is to come up with a better order
parameter related to entanglement,  rather than the entanglement
length.  There are many problems
with the entanglement length as an order parameter. The most
important one is that it  might be that the system is very
entangled, but the entanglement between two distant subsets is
zero. Such is for example a system in the state $\frac{1}{\sqrt{2}}(|0^n\ra+|1^n\ra)$,
for which any subsystem is non-entangled. Entanglement in  such
very entangled quantum systems will not be detected when looking
at sub-systems, and the entanglement length will therefore contain
no information about the actual behavior of the entanglement in
the system. Another motivation for this question is provided 
by \cite{aharonov2}, where the sizes of the clusters 
is analyzed relaxing the assumptions of nearest neighbor 
interactions. The sizes of the clusters in this case indeed 
transform from logarithmic to linear at a critical noise rate. 
However, the notion of entanglement length cannot be defined in a system 
without geometry, so it is not clear how to define an order parameter
which exhibits the phase transition in this case.

To summarize, we have discovered a phenomena of a phase transition 
in entanglement in quantum computers, and in general in quantum systems 
with local decoherence and local interactions which 
are able to generate long range entanglement in the absence of noise. 
The suggestion to explain the transition from
 quantum to classical macroscopic behavior as a phase transition
in entanglement is fundamentally different from  the
standard point of view of gradual transition, usually explained by
decoherence.
Our results have experimental implications, and raise 
a long list of open problems related to the foundations of 
quantum mechanics, as well as to quantum statistical physics.

\section{Acknowledgments}
I am most grateful to Michael Ben-Or and to Michael Nielsen.
Discussions with them inspired this work.
I am also in debt to 
Joseph Imri, David Ruelle, 
and  Wojtek Zurek, for interesting  discussions.  
I would like to thank Jennifer Chayes, Christian 
Borgs, Jeong Han Kim, David Aldous and Oded Schramm
for useful comments about classical percolation. 
Thanks to Julia Kempe, Daniel Lidar and Michael Nielsen for 
useful comments and corrections on early drafts of this paper.

\bibliographystyle{plain}

\end{document}